# Multi-path Convolutional Neural Networks Efficiently Improve Feature Extraction in Continuous Adventitious Lung Sound Detection

Fu-Shun Hsu[1,2,3], Shang-Ran Huang[3], Chien-Wen Huang[4], Chun-Chieh Chen[4], Yuan-Ren Cheng[3,5,6], and Feipei Lai[1,*]

[1] Graduate Institute of Biomedical Electronics and Bioinformatics, National Taiwan University, Taipei, Taiwan

[2] Department of Critical Care Medicine, Far Eastern Memorial Hospital, New Taipei, Taiwan

[3] Heroic Faith Medical Science Co., Ltd., Taipei, Taiwan

[4] Avalanche Computing Inc., Taipei, Taiwan

[5] Department of Life Science, College of Life Science, National Taiwan University, Taipei, Taiwan

[6] Institute of Biomedical Sciences, Academia Sinica, Taipei, Taiwan

[7] Division of Pulmonary Medicine, Far Eastern Memorial Hospital, New Taipei, Taiwan

*Corresponding author

Email: flai@csie.ntu.edu.tw

# Abstract

We previously established a large lung sound database, HF_Lung_V2 (Lung_V2). We trained convolutional-bidirectional gated recurrent unit (CNN-BiGRU) networks for detecting inhalation, exhalation, continuous adventitious sound (CAS) and discontinuous adventitious sound at the recording level on the basis of Lung_V2. However, the performance of CAS detection was poor due to many reasons, one of which is the highly diversified CAS patterns. To make the original CNN-BiGRU model learn the CAS patterns more effectively and not cause too much computing burden, three strategies involving minimal modifications of the network architecture of the CNN layers were investigated: (1) making the CNN layers a bit deeper by using the residual blocks, (2) making the CNN layers a bit wider by increasing the number of CNN kernels, and (3) separating the feature input into multiple paths (the model was denoted by Multi-path CNN-BiGRU). The performance of CAS segment and event detection were evaluated. Results showed that improvement in CAS detection was observed among all the proposed architecture-modified models. The F1 score for CAS event detection of the proposed models increased from 0.445 to 0.491–0.530, which was deemed significant. However, the Multi-path CNN-BiGRU model outperformed the other models in terms of the number of winning titles (five) in total nine evaluation metrics. In addition, the Multi-path CNN-BiGRU model did not cause extra computing burden (0.97-fold inference time) compared to the original CNN-BiGRU model. Conclusively, the Multi-path CNN layers can efficiently improve the effectiveness of feature extraction and subsequently result in better CAS detection.

**Keywords:** auscultation, continuous adventitious sound, CNN, deep learning, lung sound, GRU

# Introduction

Lung sound auscultation with a handheld stethoscope is an important technique in physical examination to evaluate a patient's pulmonary condition. The manifestation of certain types of continuous adventitious sounds (CASs), such as wheezes, stridor and rhonchi, and discontinuous adventitious sounds (DASs), such as crackles and pleural friction rubs are associated with specific pulmonary disorders [1], such as coronavirus disease 2019 (COVID-19) [2, 3] and so on. Therefore, it is crucial for a healthcare professional to accurately identify adventitious lung sound during auscultation to have a correct diagnosis.

Over the past twenty years, many computerized adventitious sound analysis algorithms have been developed on the basis of traditional machine learning [4, 5]. However, more and more studies turn to using deep learning approach to train the automated adventitious sound analysis models in recent years because of following reasons. First, unlike machine learning models, the improvement in the performance of deep learning models does not hit a plateau as the data size increases past a certain level [6]. Second, deep learning models are highly scalable [6]. You can use deeper networks or a more complicated network architecture when dealing a more complex problem. Third, deep learning can learn the features by itself and does not heavily rely on handcrafted feature engineering [6, 7]. Lastly, the learned knowledge can be transferred from one task to a new one [8].

Previously, we established two lung sound databases, HF_Lung_V1 (Lung_V1) [9] and HF_Lung_V2 (Lung_V2) [10]. Lung_V2 is an incremental expansion of open-access Lung_V1. Lung_V1 is open to the public (downloaded at https://gitlab.com/techsupportHF/HF_Lung_V1). Lung_V2 contains 14,145 15-second (s) lung sound recordings, 49,659 inhalation labels (I), 24,602 exhalation labels (E), 22,550 CAS labels (C) and 19,651 DAS labels (D). Convolutional neural network-bidirectional gated recurrent unit (CNN-BiGRU) models were used as benchmark models for detecting inhalation, exhalation, CAS and DAS in the lung sound at the recording level. The benchmark performance of inhalation, exhalation and DAS event detection (F1 scores of 0.861, 0.785 and 0.783) [10] was comparable to a previous study (F1 scores of 0.870, 0.846 and 0.721) [11]. However, the performance of CAS detection was poor (F1 score of 0.445) [10].

The possible causes of the poor model detection performance were discussed in our previous study [9]. However, compared to inhalation, exhalation and DAS detection, CAS detection faces a bigger challenge—the complexity of features—which was not described previously. CASs in the lung sounds are of three major subtypes wheezes, stridor and rhonchi [4]. CASs usually appear as high-intensity horizontal stripes on a spectrogram. However, the stripes can sometimes wiggle upward and backward (white arrow in Fig 1a) and form a curved V, W or M shape (white arrows in Fig 1b). A CAS is usually accompanied by different numbers of harmonics (clearly displayed in Fig 1a, b and c) on the spectrogram; however, the harmonics are sometimes absent (white arrows in Fig 1d–e). The fundamental frequency of a wheeze and stridor in Lung_V2 mostly ranges from 100 to 1,000 Hz. However, rhonchi was defined a continuous sound with frequency of a maximum of 200 Hz (white arrows in Fig 1c and white and blue arrows in Fig 1f) [4]. A CAS can occur in the inspiratory phase, expiratory phase or both phases (see Fig 1 and its caption) [4]. A CAS can be monophonic or polyphonic (two harmonically uncorrelated sounds start and end at roughly the same time; white arrows in Fig 1c and e) [4]. Sometimes multiple CASs take place at different regions of lung so that they are not polyphonic but they might overlap to some extent (white arrows in Fig 1d). The patterns of rhonchi are sometimes not as sharp as the ones of wheezes and stridor (white and blue arrows in Fig 1f) because a rhonchus and its harmonics cannot be clearly isolated in space-limited low-pitched range due to the limit of frequency resolution. In a nutshell, it is difficult to train a deep learning model from scratch on the basis of the diversified CAS patterns without a special network design or training strategy.

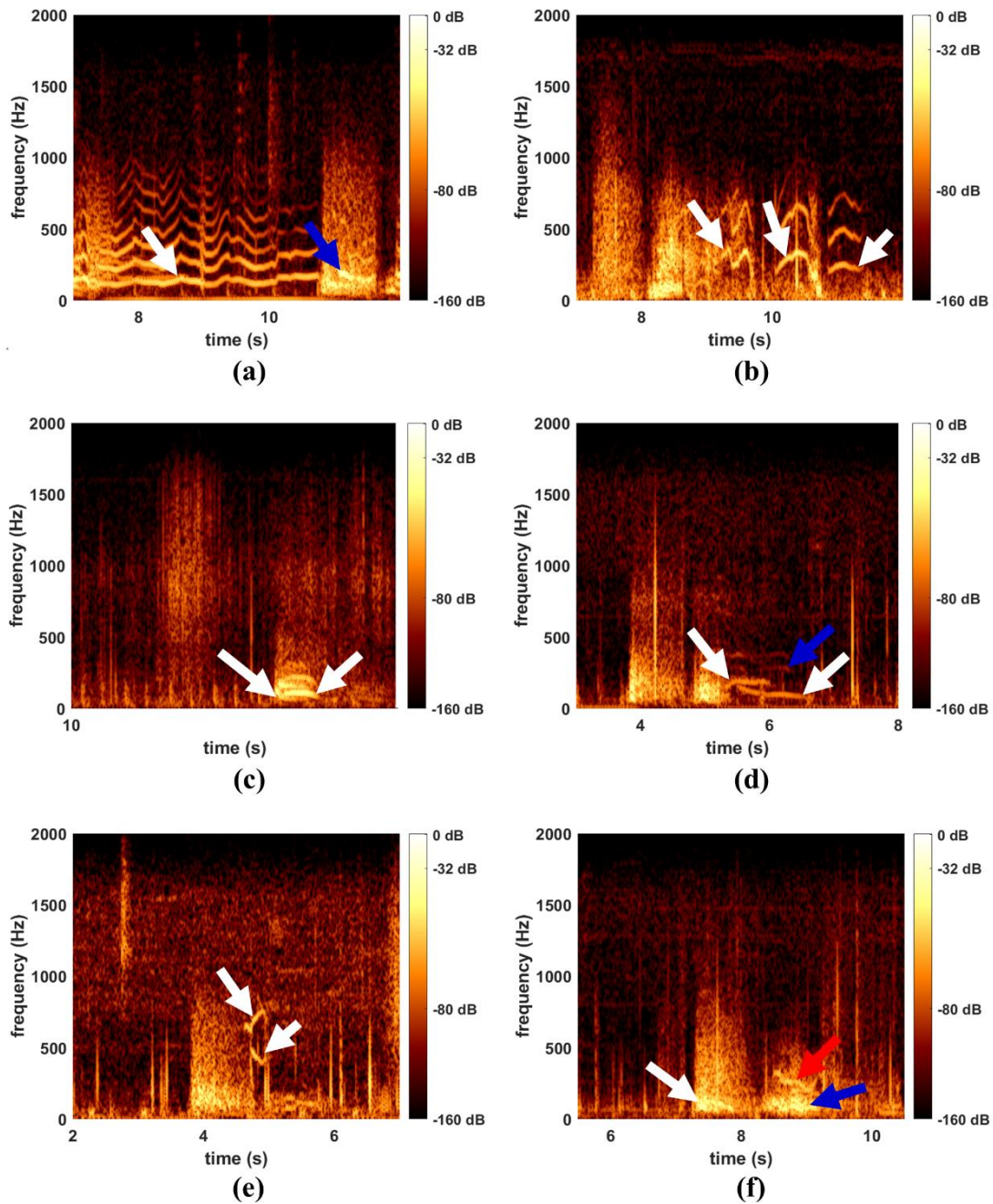

**Fig 1. Various types of CAS patterns.** (a) a long-lasting expiratory wiggling wheeze with harmonics (white arrow) and an inspiratory wheeze with harmonics (blue arrow), (b) W-shaped and inverted V-shaped expiratory wheezes with harmonics (white arrows), (c) polyphonic expiratory rhonchi (white arrows), (d) expiratory wheezes with overlapped duration (white arrows), (e) polyphonic inspiratory wheezes/stridor without harmonics (white arrows) and (f) an inspiratory rhonchus (white arrow) and an expiratory rhonchus (blue arrow) accompanied by an expiratory wheeze (red arrow).

The previously proposed CNN-BiGRU models have been run on a smartphone (Mi 9T pro, Xiaomi, Beijing, China) in the laboratory setting and real-time analysis can be reached. Because we have planned to use the respiratory sound analysis in real time on a smart device, a personal computer (PC) or a PC box, computing power is a major concern. Hence, in this study, we aimed to do minimal architectural modifications of the CNN layers in the previously proposed CNN-BiGRU model to improve CAS detection in the lung sound analysis.

# Related work

The adventitious sound analysis can be categorized into classification tasks at the segment level, event and recording levels and detection tasks at the segment and recording levels [4]. It is worthy to pursue event detection at the recording level because accurate start and end times of an event of adventitious sound can be used to compute duration and occupation rate [12, 13], which are potential outcome measures for respiratory therapy.

Many automated adventitious sound analysis methods have been proposed. However, most of the studies focused on only classifying healthy subjects and patients with respiratory diseases [14-20] and the classification of normal breathing sounds and various types of adventitious sounds [20-40]. The models in most of these studies are developed on the basis of an open-access database organized in a scientific challenge at the International Conference on Biomedical and Health Informatics (ICBHI) 2017 [41, 42]. The amount of data in ICBHI database is smaller than that in Lung_V2.

So far, only Messner et al. [11] have reported adventitious lung sound detection at the recording level based on deep learning. However, they only developed a BiGRU model for crackle detection, which achieved an F1 score of approximately 72%. To our best knowledge, as of writing this paper, no one has reported computerized CAS detection at the recording level based on deep learning.

## Materials and Methods

### Database: Lung_V2

The CAS sound recordings and labels used in this study were from the lung sound database, Lung_V2 [10], developed by Heroic Faith Medical Science Co. Ltd., Taipei, Taiwan. The information about the recording devices and the participants from whom the lung sound was acquired, and the protocol of sound recording and labeling can be found in our previous studies [9, 10].

Lung_V2 contains 14,145 15-second (s) lung sound recordings. In these recordings, the start and end times of all types of CASs are labeled. A total of 14,128 wheeze labels (W), 917 stridor labels (S) and 7,505 rhonchus labels (R), which collectively form 22,550 CAS labels (C), are included in the Lung_V2. The sound recordings and corresponding labels have already been assigned to either training dataset or test dataset in the Lung_V2. The statistics of sound recordings and CAS labels are tabulated in Table 1.

**Table 1: Statistics of CAS labels in the training and test datasets of Lung_V2.**

| Labels | Training dataset | Test dataset | Total |
|---|---|---|---|
| No. of 15-sec. recordings | 10742 | 3403 | 14145 |
| Total duration (min) | 2685.50 | 850.75 | 3536.25 |
| No. of C/W/S/R | 18353/12442/869/5042 | 4197/1686/48/2463 | 22550/14128/917/7505 |
| Duration of C/W/S/R (min) | 255.77/179.11/12.36/64.29 | 52.45/23.19/0.48/28.77 | 308.21/202.31/12.84/93.06 |
| Mean duration of C/W/S/R (s) | 0.84/0.86/0.85/0.77 | 0.75/0.83/0.60/0.71 | 0.82/0.86/0.84/0.74 |

C: CAS labels, W: wheeze labels, S: stridor labels, R: rhonchus labels.

## Analysis pipeline

CASs consist of three major subtypes, namely wheezes, stridor and rhonchi. However, a clear-cut borderline between various subtypes of CASs is lacking. The terms of rhonchi and low-pitched wheezes are sometimes used interchangeably [43], and stridor is occasionally classified as a subtype of wheezes [44]. Thus, we trained an overall CAS detection model but not respective wheeze, stridor and rhonchus detection models in our previous studies, and we intended to follow this approach in the present study.

The analysis pipeline is displayed in Fig 2. The whole analysis consisted of a preprocessing step, a prediction step, and a postprocessing step.

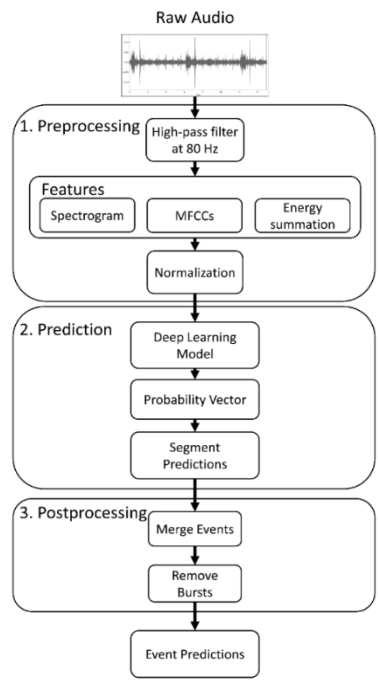

**Fig 2. Analysis pipeline.**

During the preprocessing, we derived several features from the raw signal of the 15-s lung sound. A high-pass Butterworth filter with cut-off frequency of 80 Hz was first applied to the 15-s raw signal. Then, short-time Fourier transform [9, 45] (window type: Hanning, window size: 256, overlap ratio: 0.75, zero-padding: none) was used to compute the spectrogram of the filtered signal, which produced a 129 × 938 matrix.

The value of 129 is the number of frequency bins and the value of 938 is the number of time frames (time segments). Additionally, the mel frequency cepstral coefficients (MFCCs) [9], including 20 static coefficients, 20 delta coefficients, and 20 acceleration coefficients, were also generated for each of the 938 time segments from the spectrogram. Finally, the energy summation of four frequency bands in each time segment, namely, 0-250, 250-500, 500-1,000, and 0-2,000 Hz, of the spectrogram were calculated, which yielded four $1 \times 938$ vectors. Normalization was then applied to the spectrogram, 20 static MFCCs, 20 delta MFCCs, 20 acceleration MFCCs and each vector of the energy summation. As a result, the 15-s signal was broken down into 938 time segments, each of which had a total of 193 features (129 from the spectrogram, 60 from the MFCCs and 4 from the energy summation). The preprocessed data were then fed into the deep learning models for the CAS detection.

In the prediction step, the deep learning model made an inference on the basis of input features and gave a prediction result, which was a $1 \times k$ probability vector. The value of $k$ is determined by the architecture of the deep learning models. The value in each element of the vector represented the probability of CAS being present in the corresponding time segment. Then, thresholding was applied to probability vector to give the final segment prediction results. The elements in the vector were set as 1 if the output was greater than or equal to a thresholding value; otherwise, the element was set as 0. The value of 1 indicated CAS was detected in the corresponding time segment.

After we had the results of segment prediction, the vector was sent to postprocessing for merging neighboring segments and removing burst events to generate the results of event detection. In the merging step, we first regarded the connected segments as independent events. Then, we checked the continuity of all the pairs of two neighboring independent events iteratively and decided whether a pair of events can be merged to a single one. If the interval between the $j$th and $i$th events was smaller than $T$ s, then we computed the difference in frequency between their energy peaks ($|p_j - p_i|$). Subsequently, if the frequency difference was below a given threshold $P$, the two neighboring events were merged into one. In this study, $T$ was set to 0.5 s, and $P$ was set to 25 Hz. Following the merging step, we further examined if a burst event existed. If the duration of a remained event was shorter than 0.05 s, the event was deleted.

## Modified Deep Learning Models

According to a previous review paper [46], the significant improvement of CNN performance was achieved through architectural innovations. The strategies used to modify the CNN architecture can generally be classified into seven categories, spatial exploitation, depth, multi-path, width, feature-map exploitation, channel boosting, and attention [46]. In this study, we modified the architecture of CNN layers of the CNN-BiGRU model (Fig 3) used in our previous studies on the basis of three strategies, making the CNN layers (1) deeper, (2) wider and (3) receive multiple feature inputs in different paths.

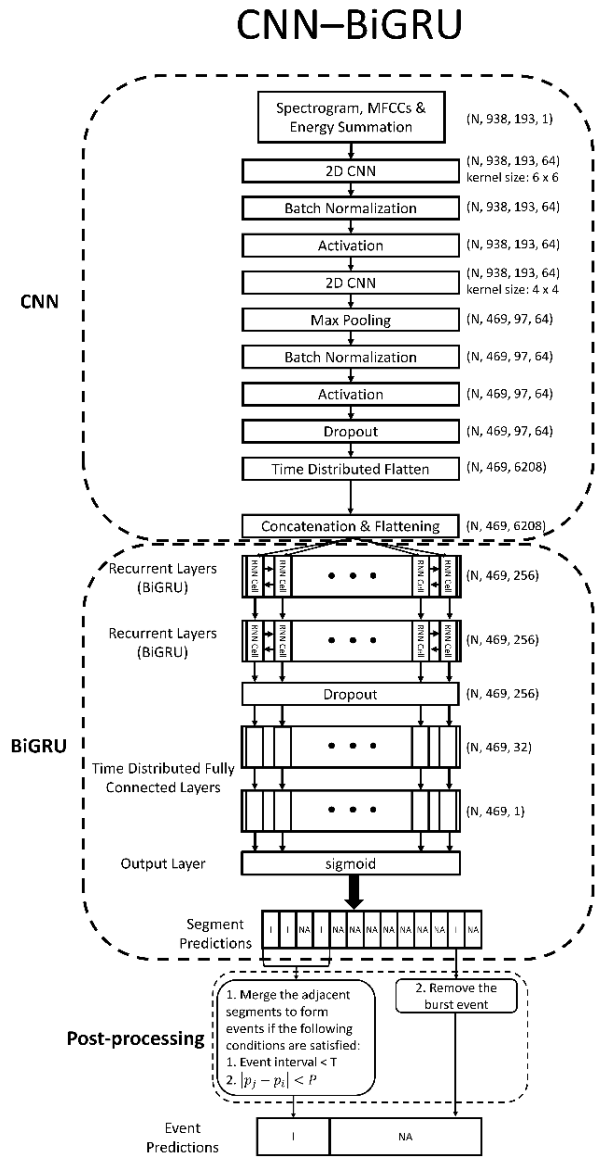

**Fig 3. The architecture of CNN-BiGRU model used in the previous studies [9, 10].**

Firstly, we investigated the performance of deeper CNN layers. Because the deep residual network (ResNet) was reported to be capable of solving the vanishing gradient problem encountered in training deep CNN networks [24], we did not just add more CNN layers into the original CNN-BiGRU model but used residual block (RB; Fig 4a) instead. The RB contains two 2D CNN layers with a kernel size of 3 × 3, a batch normalization layer, two activation layers, and a shortcut connection (Fig 4a). We added one RB followed by a max pooling layer, a dropout layer and a flattening layer to replace the CNN component in the original CNN-BiGRU model and formed a RB1-BiGRU model (Fig 4b). Similarly, a RB2-BiGRU model (Fig 4c) was created by using two consecutive RBs followed by a max pooling layer, a dropout layer and a flattening layer. Note that the CNN kernel size in the RB was all 3 × 3, while the kernel size was 6 × 6 in the first CNN layer and 4 × 4 in the second CNN layer in the original model. Both the RB1-BiGRU and CNN-BiGRU models had two CNN layers in the architecture, therefore the RB1-BiGRU model was not considered substantially deeper than the original CNN-BiGRU model. However, the RB2-BiGRU model had four CNN layers in the architecture, it was a deeper network compared to the CNN-BiGRU model.

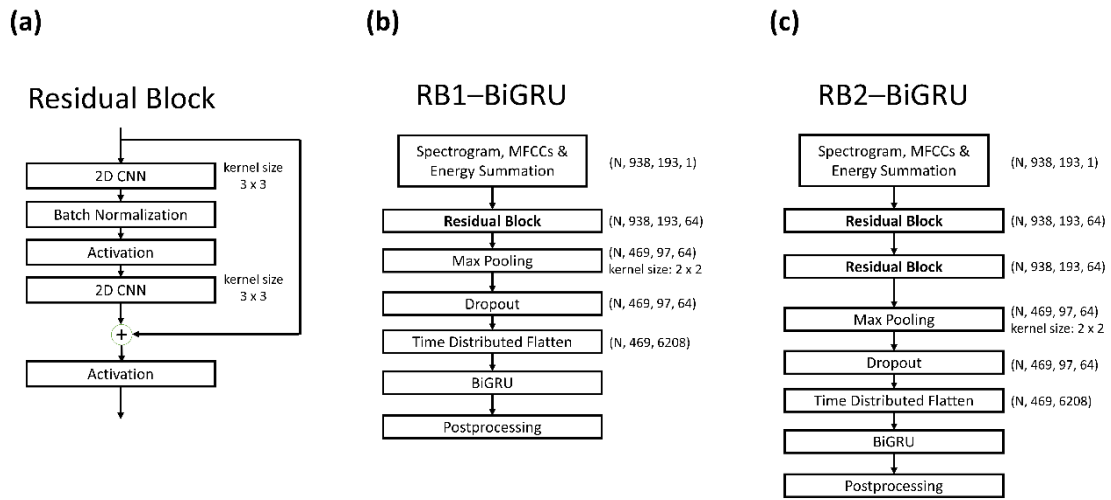

**Fig 4. Architectures of (a) the residual block (RB), (b) the model with one RB (RB1-BiGRU) and (c) the model with two RBs (RB2-BiGRU).**

Secondly, we increased the number of CNN kernels in the original CNN-BiGRU model to either 96 or 128 to make two new models, denoted by CNN96-BiGRU (Fig 5a) and CNN128-BiGRU (Fig 5b). Note that the number of CNN kernels in the CNN-BiGRU model was 64. Since more CNN kernels meant more extracted features, we viewed CNN96-BiGRU and CNN128-BiGRU as wider networks.

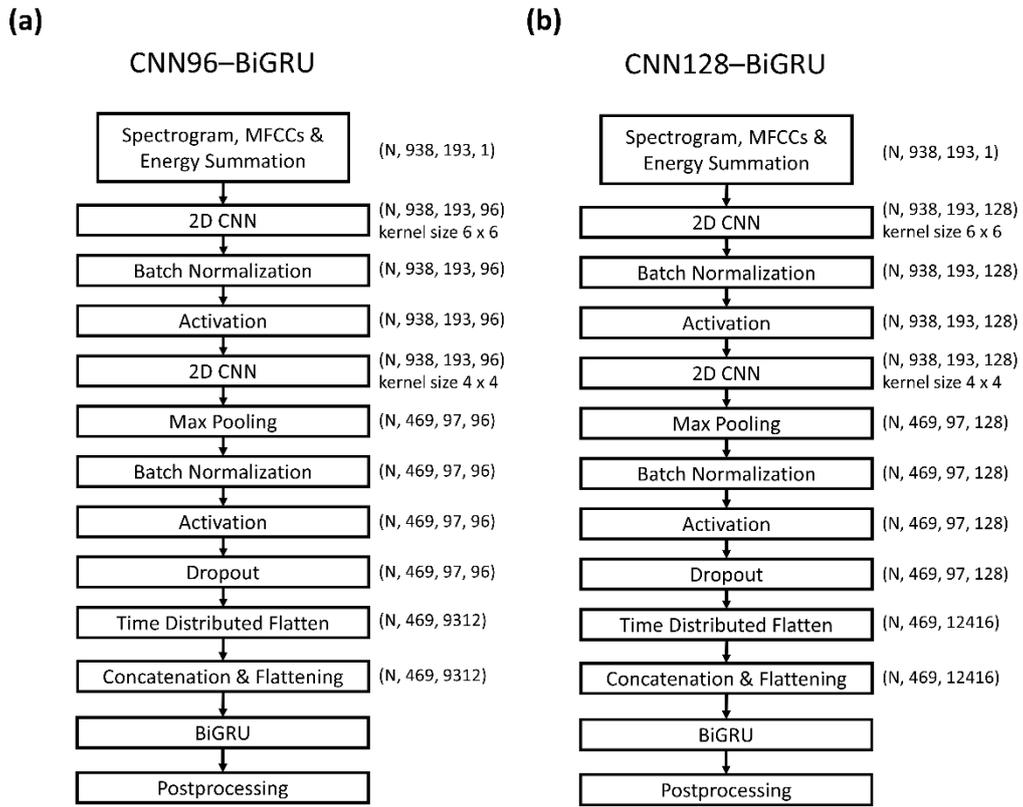

Fig 5. Architectures of (a) the model with CNN kernel number of 96 (CNN96-BiGRU), and (b) the model with CNN kernel number of 128 (CNN128-BiGRU).

Thirdly, the concept of multi-path CNN networks [47] was adopted. We modified the original CNN-BiGRU model to have two feature inputs in two different paths (Fig 6) and named it as a Multi-path CNN-BiGRU model. In this model, spectrogram was processed in one path, and the concatenation of MFCCs and energy summation were processed in the other path. The formation and the order of the deep learning layers were the same in the two paths of the Multi-path CNN -BiGRU model and in the original CNN-BiGRU model.

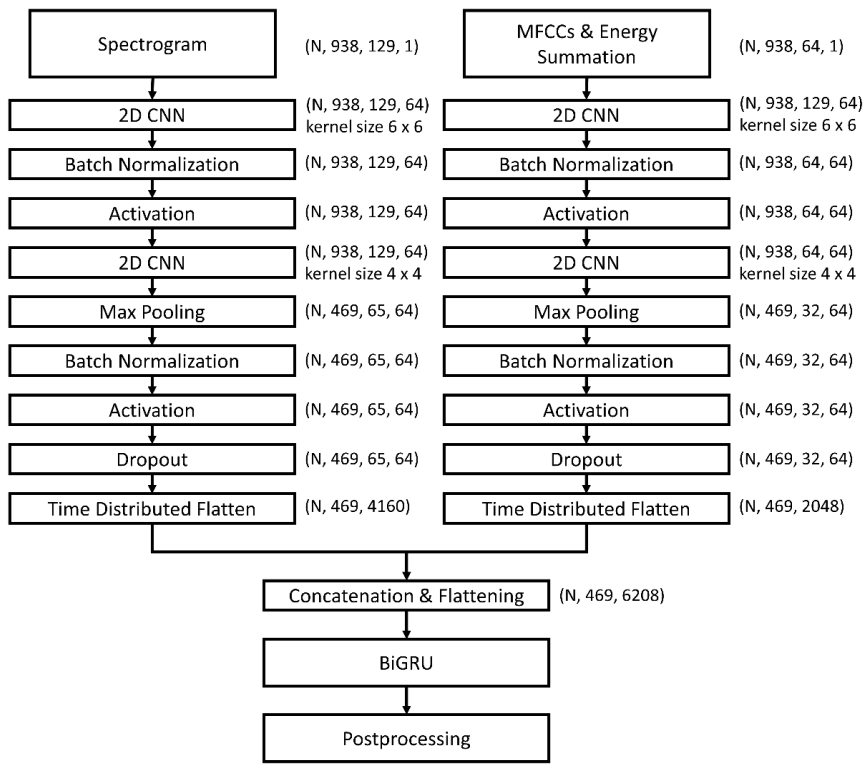

**Fig 6. Architecture of the Multi-path CNN-BiGRU model.**

## Training parameters and environment

Fivefold cross-validation was adopted in the training dataset to train and validate the models. We used Adam optimizer and we set the initial learning rate to 0.0001 with a step decay (0.2×) when the validation loss did not decrease for 10 epochs. The learning process stopped when no improvement occurred over 50 consecutive epochs.
Note that only the recordings contained at least a CAS label were used to train the
model and evaluate the model performance.

We trained the models on a server (OS: Ubuntu 18.04; CPU: Intel(R) Xeon(R) Gold 6154 @3.00 GHz; RAM: 90 GB) provided by the National Center for High-Performance Computing in Taiwan [Taiwan Computing Cloud (TWCC)]. We used TensorFlow 2.10 as the framework to build the deep learning models. CUDA 10 and CuDNN 7 programs were used to run GPU acceleration on a NVIDIA Titan V100 card.

## Performance evaluation

Fig 7 illustrates how we evaluated the performance of CAS detection at the recording level. After we had ground-truth labels in the 15-s recordings (red horizontal bars in Fig 7a), we can use them to create the ground-truth time segments (red vertical bars in Fig 7b). By comparing the ground-truth time segments (red vertical bars in Fig 7b) with the results of segment prediction (blue vertical bars in Fig 7c), we could define true positive (TP; orange vertical bars in Fig 7d), true negative (TN; green vertical bars in Fig 7d), false positive (FP; black vertical bars in Fig 7d), and false negative (FN; yellow vertical bars in Fig 7d) time segments, which were used to evaluate the performance of CAS segment detection of the models.

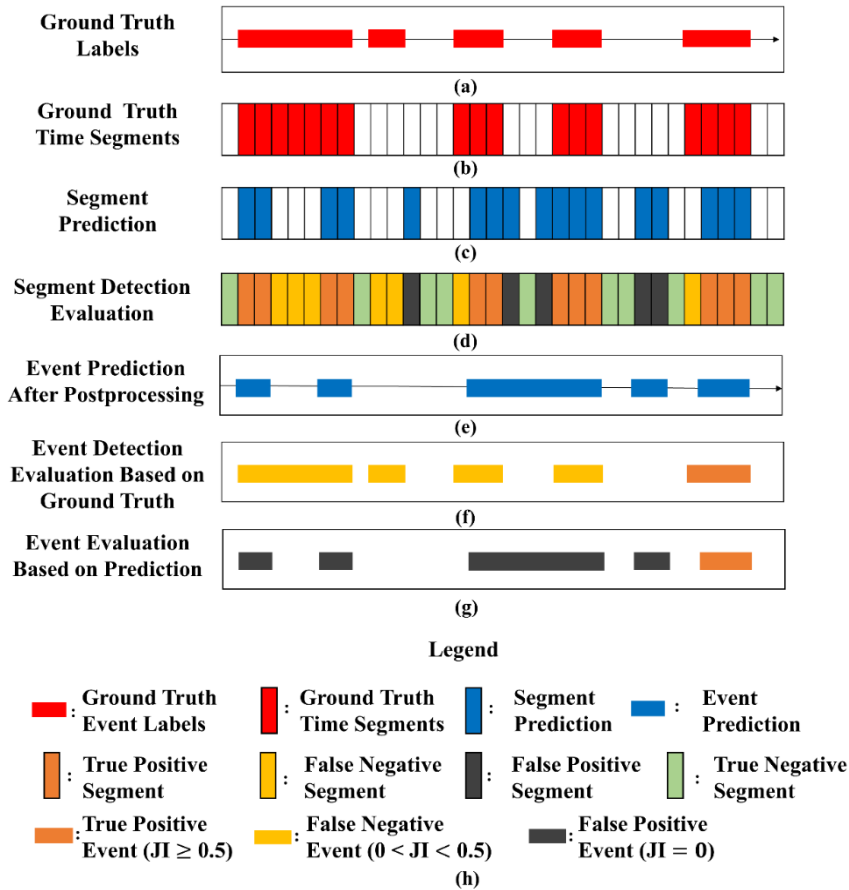

**Fig 7. Illustration of segment and event detection evaluation.** (a) Ground-truth event labels, (b) ground-truth time segments, (c) segment prediction, (d) segment detection evaluation, (e) event prediction after postprocessing, (f) event detection evaluation based on ground-truth event labels, (g) event evaluation based on event prediction, and (h) legend. JI: Jaccard index.

As for the event detection, after the postprocessing was applied to the segment prediction results, we can have the event prediction results (blue horizontal bars in Fig 7e). Then, we used Jaccard index (JI) to determine whether the models correctly detected an CAS event. The JI was defined as the value of the intersection of two sets (A and B) divided by the union of them:

$$JI = \frac{A \cap B}{A \cup B},$$

where the set A was the period of a ground-truth label, and the set B was the period of a predicted event. We first used the ground-truth labels as a reference and checked whether every ground-truth label had a predicted event matched to it (JI $\geq$ 0.5). If not, we counted it as an FN event (the yellow horizontal bars in Fig 7f); otherwise, it was a TP event (the orange horizontal bar in Fig 7f). Then, conversely, we used the event prediction results as a reference in this time. We checked whether we could find a matched ground-truth label for every predicted event (JI $\geq$ 0.5). If not, the predicted event was regarded as an FP event (the black horizontal bars in Fig 7g); otherwise, it was a TP event (the orange horizontal bar in Fig 7g). A TN event cannot be defined in the event detection. Note that the TP events were double counted (the orange horizontal bars in Fig 7f and g) as the ground-truth labels and the event prediction results were used as a reference by turns. Therefore, we counted a pair of TP events as a single TP event but summed up all the FN (Fig 7f) and FP events (Fig 7g) to form the final event detection results.

The performance of the segment detection was measured by the accuracy (ACC), predictive positive value (PPV), sensitivity (SEN), specificity (SPE), F1 score, and area under the receiver operating characteristic (ROC) curve (AUC), whereas the event detection was evaluated with PPV, SEN, and F1 score. The thresholding value rendering the best ACC of segment detection was used to calculate PPV, SEN, SPE and F1 score. Because we pursued the event detection more than segment detection, F1 score was used as the major index.

## Results

The performance of CAS detection based on the proposed models are tabulated in Table 2. The CNN-BiGRU model had the best SPE (0.963) in the segment detection. The CNN96-BiGRU model had the best ACC (0.885) and AUC (0.916) in the segment detection. The CNN96-96-BiGRU and CNN128-BiGRU models had the best PPV (0.687) among all the models. The Multi-path CNN-BiGRU model had the best SEN (0.505) and F1-score (0.575) in the segment detection and the best PPV (0.498), SEN (0.432) and F1-score (0.530) in the event detection.

The number of trainable parameters of each model and the ratio of inference time cost of each model to the original CNN-BiGRU model were also listed in the Table 2. The numbers of parameters in the RB1-BiGRU (5.21 M), RB2-BiGRU (5.28 M) and Multi-path CNN-BiGRU (5.31 M) models did not have a big difference compared to the CNN-BiGRU model (5.24 M); thus, the time cost to complete an inference was comparable between these models. However, the CNN96-BiGRU and CNN128-BiGRU models took 1.57-fold and 2.44-fold time to complete an inference as the number of parameters increased from 5.24 M to 7.71 M and 12.21 M, respectively.

**Table 2. Performance of CAS detection of all the models.**

| Model | No. of Parameters | Inference Time Cost | Segment Detection | | | | | | Event Detection | | |
|---|---|---|---|---|---|---|---|---|---|---|---|
| | | | ACC | PPV | SEN | SPE | F1 | AUC | PPV | SEN | F1 |
| CNN–BiGRU | 5,240,513 | 1.00× | 0.877 | 0.671 | 0.411 | **0.963** | 0.509 | 0.896 | 0.399 | 0.347 | 0.445 |
| RB1–BiGRU | 5,210,113 | 0.82× | 0.880 | 0.676 | 0.440 | 0.961 | 0.532 | 0.910 | 0.428 | 0.406 | 0.491 |
| RB2–BiGRU | 5,284,225 | 1.18× | 0.881 | 0.677 | 0.454 | 0.960 | 0.543 | 0.909 | 0.424 | 0.414 | 0.500 |
| CNN96–BiGRU | 7,707,649 | 1.57× | **0.885** | 0.687 | 0.485 | 0.959 | 0.568 | **0.916** | 0.461 | 0.425 | 0.520 |
| CNN128–BiGRU | 10,207,553 | 2.44× | 0.882 | **0.687** | 0.453 | 0.961 | 0.545 | 0.910 | 0.436 | 0.408 | 0.503 |
| Multi-path CNN–BiGRU | 5,308,737 | 0.97× | 0.884 | 0.671 | **0.505** | 0.954 | **0.575** | 0.914 | **0.498** | **0.432** | **0.530** |

ACC: accuracy, PPV: positive predictive value, SEN: sensitivity, SPE: specificity, F1: F1 score, AUC: area under the receiver operating characteristic curves.

The ROC curves of all the models are presented in Fig 8. In the legend, the number in the first parenthesis following the model name was the number of the trainable parameters and the AUC was listed in the second parenthesis. All the proposed models performed better than the previous CNN-BiGRU model. The CNN96-BiGRU model overall exhibited the best performance over a wide range of thresholding values (green dashed line in Fig 8 with AUC of 0.916). The Multi-path CNN-BiGRU model had the second best performance (red solid line in Fig 8 with AUC of 0.914).

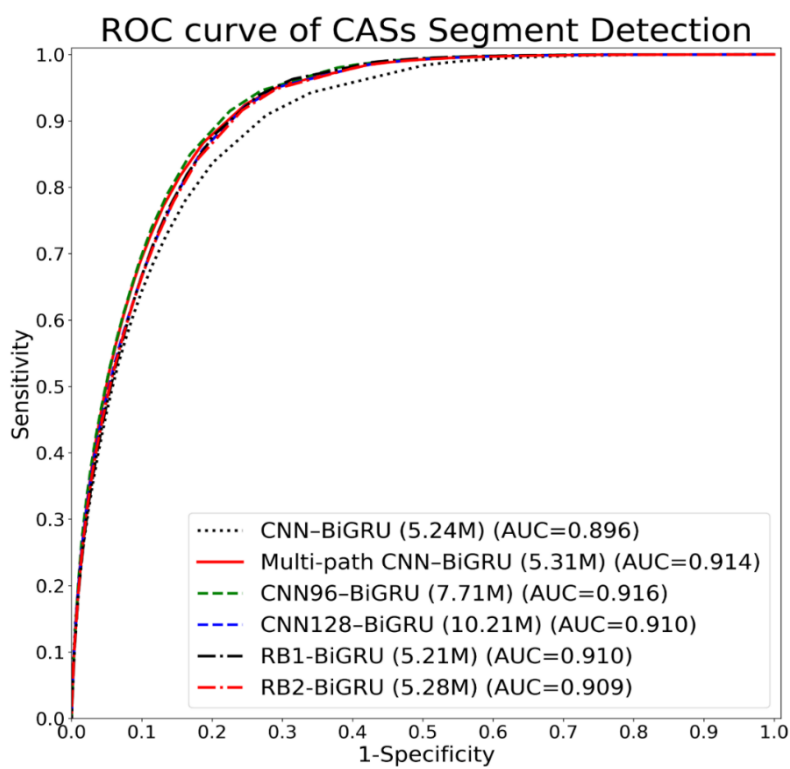

**Fig 8. The ROC curves and AUC values of CAS in segment detection among variant models.**

## Discussions

The results showed that the performance of CAS detection was significantly improved in all the architecture modified models. The F1 scores of the proposed models increased from 0.445 to 0.491–0.530 (Table 2) and the ROC curves of them (red solid line, green dashed line, blue dashed line, black dot-and-dash line and red dot-and-dash lines in Fig 8) had some distance away from the curve of the CNN-BiGRU model (black dotted line in Fig 8). Because we do not change the BiGRU layers in the models, it implies that the improvement in CAS detection directly results from the effectiveness of feature learning in the CNN layers.

Among all the models, the Multi-path CNN-BiGRU model is deemed the best one because it had the best performance in five evaluation metrics, including F1 score. Furthermore, compared to the original CNN-BiGRU model, the Multi-path CNN-BiGRU model does not take extra time to execute CAS detection.

Although a significant improvement is observed in the proposed models, the performance of CAS detection is still far from ideal. To further improve the accuracy of CAS detection, several approaches can be used. First, find a way to mitigate the data imbalance problem [48]. Second, data augmentation methods, such as vocal track length perturbations [49], noise addition [50], time stretching [50], frequency and time masking [51], can be applied. Third, more basic features [52], higher order spectrum.[53, 54] and feature engineering methods, such as harmonic percussive source separation [55], can be explored. Third, the labeling can be expanded to two-dimensional bounding-boxes and facilitate the use of the state-of-the-art object detection deep learning models [56]. Lastly, more accurate ground-truth labels can be established.

The major cause of poor CAS detection is believed to be the quality of labeling. We are currently reviewing the labels and build up a more accurate ground-truth labels. The corrected labels in Lung_V1 will be updated and released (https://gitlab.com/techsupportHF/HF_Lung_V1) [9] though the incremental labels in Lung_V2 will not be open to the public.

## Conclusion

In this study, we show that multi-path CNN layers can improve the CAS feature extraction at no extra time cost and it subsequently help the following BiGRU networks better detect pattern-diversified CASs. However, the performance of the CAS detection model still needs a significant improvement before it can be put into real world application. Nevertheless, multi-path strategy is worth consideration when developing a more accurate CAS detection model in the future.

# Acknowledgment

The authors thank the employees of Heroic Faith Medical Science Co. Ltd. who have ever contributed to this study. The author would like to acknowledge the National Center for High-Performance Computing (TWCC) in providing computing resources to facilitate this research.